\documentclass[reprint,amsmath,amssymb,aps,pra,showpacs]{revtex4-1}
\usepackage[latin9]{inputenc}
\usepackage{xcolor}
\usepackage{amstext}
\usepackage{graphicx}
\PassOptionsToPackage{normalem}{ulem}
\usepackage{ulem}

\makeatletter

\providecolor{lyxadded}{rgb}{0,0,1}
\providecolor{lyxdeleted}{rgb}{1,0,0}

\usepackage{dcolumn}
\usepackage{bm}
\usepackage[toc,page]{appendix}

\makeatother

\begin{document}

\title{Many-body localization in Ising models with random long-range interactions}

\author{Haoyuan Li}

\thanks{Chris.Lee.China.1993@gmail.com, present address: Department of Physics,
Stanford University, Stanford, California 94305, USA}

\affiliation{Department of Physics, Tsinghua University, Beijing 100084, China}

\author{Jia Wang}

\affiliation{Centre for Quantum and Optical Science, Swinburne University of Technology,
Melbourne 3122, Australia}

\author{Xia-Ji Liu}

\affiliation{Centre for Quantum and Optical Science, Swinburne University of Technology,
Melbourne 3122, Australia}

\author{Hui Hu}

\thanks{hhu@swin.edu.au}

\affiliation{Centre for Quantum and Optical Science, Swinburne University of Technology,
Melbourne 3122, Australia}

\date{\today}
\begin{abstract}
We theoretically investigate the many-body localization phase transition
in a one-dimensional Ising spin chain with random long-range spin-spin
interactions, $V_{ij}\propto\left|i-j\right|^{-\alpha}$, where the
exponent of the interaction range $\alpha$ can be tuned from zero
to infinitely large. By using exact diagonalization, we calculate
the half-chain entanglement entropy and the energy spectral statistics
and use them to characterize the phase transition towards the many-body
localization phase at infinite temperature and at sufficiently large
disorder strength. We perform finite-size scaling to extract the critical
disorder strength and the critical exponent of the divergent localization
length. With increasing $\alpha$, the critical exponent experiences
a sharp increase at about $\alpha=1$ and then gradually decreases
to a value found earlier in a disordered short-ranged interacting
spin chain. For $\alpha<1$, we find that the system is mostly localized
and the increase in the disorder strength may drive a transition between
two many-body localized phases. In contrast, for $\alpha>1$, the
transition is from a thermalized phase to the many-body localization
phase. Our predictions could be experimentally tested with ion-trap
quantum emulator with programmable random long-range interactions,
or with randomly distributed Rydberg atoms or polar molecules in lattices.
\end{abstract}

\maketitle

\section{Introduction}

Many-body localization (MBL) phase transition \cite{Basko2006,Nandkishore2015,Altman2015}
is an interesting phenomenon and has caught huge attention of physicists
in various perspectives. From a theoretical point of view, this localization
phenomenon addresses one of the most fundamental problems in statistical
mechanics: How does a generic quantum many-body system evolve, when
it is absolutely isolated from environment? The fundamental Schrödinger
equation governing the dynamics of a quantum system is a linear equation,
which renders the traditional explanations of thermalization, i.e.,
via ergodicity and equilibrium states, inapplicable \cite{Gibbs1878,Moore2015}.
Besides attempting to adapt the original concepts and methods to new
situations \cite{Neumann1932a,Neumann1932b}, during the last several
decades, another extensively explored approach is to establish an
alternative hypothesis. The eigenstate thermalization hypothesis (ETH)
\cite{Srednicki1994,Rigol2012,Zhao2015} is one of the most successful
hypotheses. It asserts that in a generic quantum many-body system,
for nearly every highly excited eigenstate, averaging thermodynamic
observable over that eigenstate essentially leads to the classical
equilibrium value. This hyperthesis results in significant implications
and profound insights into the thermalization process and was proved
to be equivalent to an crucial assumption, i.e., the von Neumann's
quantum ergodic theorem \cite{Neumann1932a,Neumann1932b}. However,
to some extent, one of the most important prospects of this hypothesis
is to provide a manipulable criterion to characterize the breakdown
of equilibrium statistical mechanics.

Anderson localization (AL) phase transition in non-interacting systems
\cite{Anderson1958,Evers2008} proves that it is an over-simplified
picture in ETH to assume that equilibrium states always exist. As
a natural analogy of AL phase transition in interacting systems, MBL
phase transition pushes the inquiry of the boundary of that picture
to a new realm: Can strong enough quenched disorder localizes a generic
quantum many-body system, or at least breaks the eigenstate thermalization
hypothesis? Exploring such profound questions will lead to a deeper
understanding of the non-trivial role of interactions and disorder
in evolutions of many-body systems and fundamental concepts of statistical
mechanics, such as temperature, ergodicity and extensiveness in quantum
statistical mechanics and phase transition.

This topic has attracted huge interest ever since Anderson's seminal
paper \cite{Anderson1958} on AL phase transition. Anderson himself
already showed interest in extending the localization phase transition
concept to systems with interactions. He and many successors have
tried to prove the existence of the localized phase in interacting
models. (For recent reviews, see for example, Refs. \cite{Nandkishore2015}
and \cite{Altman2015}.) Thanks to their efforts, many exotic features
of the MBL phase and phase transition have been discovered: the existence
of the mobility edge \cite{Luitz2015,Baygan2015,Li2015}, vanishing
of conductivity at non-zero temperature in the MBL phase \cite{Berkelbach2010,Pekker2014,Agarwal2015},
existence even with weak connection to thermal bath \cite{Nandkishore2014},
sustaining long-range order in situations where equilibrated systems
would not \cite{Huse2013,Bauer2013,Bahri2013,Chandran2014,Vosk2014},
half chain entanglement entropy of (highly excited) eigenstates obeying
an area-law \cite{Znidaric2008,Bauer2013,Kjall2014}, possessing an
extensive number of local integral motions \cite{Serbyn2013a} and
universal slow growth of entanglement of eigenstates after a sudden
quench \cite{Znidaric2008,Bardarson2012,Vosk2013,Serbyn2013b,Vosk2014}.
These features are also taken as indicators for the MBL phase.

From an experimental (and engineering) point of view, MBL phase and
phase transition are also of great interest and importance. Localization
implies that the information stored through local freedom within the
initial state will remain within local freedoms \cite{Gogolin2011}.
This may help to develop future quantum devices for quantum information
process \cite{Nandkishore2015,Altman2015}. In addition, the rapid
progress of cold atom experiments over the past two decades has immensely
expanded our ability to control the experimental conditions for studying
quantum physics \cite{Bloch2008}. Especially in a dilute quantum
gas with or without optical or magnetic lattice, we can induce different
forms and tune the strength of interactions to a large extent at our
will \cite{Chin2010}. This makes the direct observation and simulation
of quantum many-body systems possible \cite{Bloch2012}. Experimental
exploration is now one of the most promising approaches towards this
phenomenon, with several experiments successfully implemented, including
both neutral atom gas systems \cite{Schreiber2015,Bordia2016,Choi2016}
and ion traps \cite{Smith2015}. Both short-range interaction and
long-range interaction have been realized. On the other hand, because
of the quantum nature of many-body systems, theoretical study of MBL
phase transition is in general quite challenging. 

At present, in the extensive literature, three kinds of theoretical
approaches are most popular to tackle the MBL challenge: resonance
analysis \cite{Ros2014,Yao2014,Laumann2014,Hauke2015}, renormalization
group methods \cite{Vosk2013,Hauke2015,Karrasch2015,Vosk2015,Potter2015,Zhao2015,Lim2016}
and numerical exact diagonalization \cite{Oganesyan2007,Pal2010,Bardarson2012,Serbyn2013a,Luca2013,Luitz2015,Li2015,Agarwal2015,Baygan2015,Serbyn2016}.
Each method has its own pros and cons. Resonance analysis and renormalization
group can be applied to relatively large systems. However, it is challenging
to justify the validity of these approaches. On the other hand, numerical
exact diagonalization is reliable but can only handle rather small
systems.

\begin{figure}
\centering{}\includegraphics[width=0.48\textwidth]{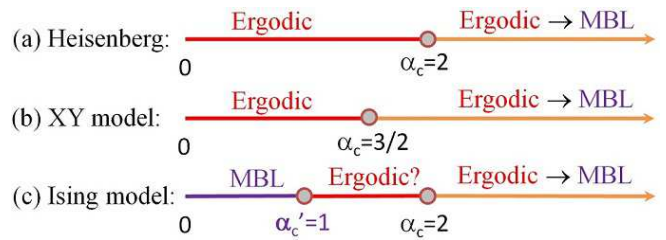} \caption{(color online). Proposed phase diagrams of 1D long-range interacting
spin models under random magnetic fields in the thermodynamic limit:
(a) the Heisenberg model \cite{Yao2014,Burin2015a}, (b) the XY model
\cite{Burin2015b} and (c) the Ising model \cite{Burin2015a,Hauke2015}.
For all three spin models, it was predicted that the system is de-localized
and ergodic when $\alpha<\alpha_{c}$; while for $\alpha>\alpha_{c}$
the system experiences a transition from the de-localized state to
the many-body localization state with increasing disorder strength.
For the Ising model in (c), an additional many-body localization phase
is suggested to appear at any nonzero disorder strength when $0\leqslant\alpha\leqslant\alpha_{c}'=1$
\cite{Hauke2015}. \label{PhaseDiagramPreviousWorks}}
\end{figure}

All of these approaches has been applied for systems with short-range
interactions, such as Heisenberg model \cite{Znidaric2008,Luitz2015,Baygan2015,Agarwal2015}
and Ising model \cite{Canovi2011,Kjall2014,Vosk2014}. Nevertheless,
there have been only few numerical studies on systems with long-range
interactions that might be of more interest for experimentalists,
including trapped ions, Rydberg atoms, and polar molecules with dipole-dipole
interactions \cite{Yao2014,Pino2014,Burin2015a,Burin2015b,Hauke2015,Smith2015,Wu2016}
where the MBL physics in these systems is less understood. Using long-range
interacting spin chains under random transverse fields as an example,
it has been recently suggested that the spin-spin interaction in the
form of $V_{ij}\propto\left|i-j\right|^{-\alpha}$ always leads to
the breakdown of the localization in the thermodynamic limit when
$\alpha<\alpha_{c}$, where $\alpha_{c}$ is a threshold interaction
exponent that depends on the dimensionality $d$ of the system. By
constructing hierarchical spin resonances \cite{Yao2014}, one obtains
$\alpha_{c}=2d$ for the Heisenberg and Ising models \cite{Burin2015a}
and $\alpha_{c}=3d/2$ for the XY model \cite{Burin2015b}. However,
this picture of the so-called interaction-assisted delocalization
is not unambiguously settled. For a one-dimensional (1D) transverse-field
Ising chain, an additional many-body localization phase has been shown
by Hauke and Heyl to occur at $0\leqslant\alpha\leqslant1$, using
both numerical simulation and non-equilibrium dynamical renormalization
group analysis \cite{Hauke2015}. The various proposed phase diagrams
of 1D disordered long-range spin chains are briefly summarized in
Fig. \ref{PhaseDiagramPreviousWorks}.

\begin{figure}
\centering{}\includegraphics[width=0.48\textwidth]{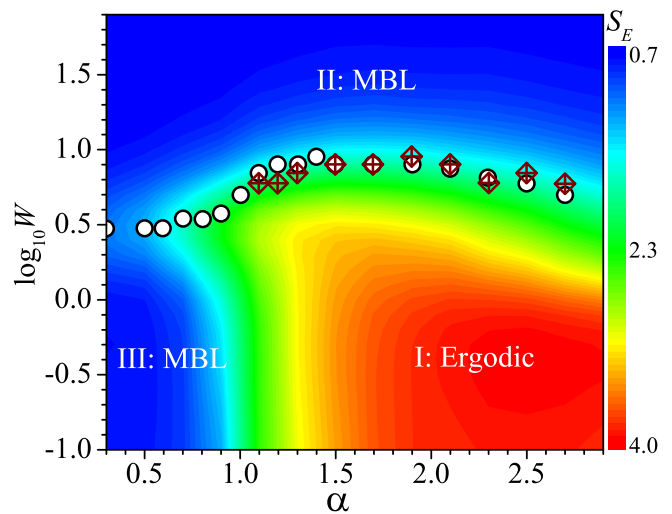} \caption{(color online). Contour plot of the half-chain entanglement entropy
of a $L=14$ Ising spin chain with random long-range interactions
(see the Hamiltonian Eq. (\ref{(H)})) at infinite temperature, as
functions of the interaction exponent $\alpha$ and the disorder strength
$W$ (in the logarithmic scale). A thermalized phase (I) and two many-body
localization phases (II and III) are clearly visible. The symbols
show the phase boundary determined from the finite scaling of the
entanglement entropy (circles) and of the energy spectral statistics
(diamonds). Here, we take a uniform transverse $B=0.6J$. \label{PhaseDiagramL14}}
\end{figure}

This work aims to shed some light on settling the above-mentioned
debate, by performing extensive numerical calculations via exact diagonalization
for a specific Ising model with disordered long-range interactions.
Our main results are summarized in Fig. \ref{PhaseDiagramL14}, which
reports the half-chain entanglement entropy ($S_{E}$) of our Ising
model with length $L=14$. One may easily identify an ergodic, thermalized
phase and two MBL phases. The phase boundary between ergodic and MBL
phases has been been determined by a finite-size scaling analysis.
Our results strongly support the conclusion by Hauke and Heyl that
there is an MBL phase at any nonzero disorder strength in the long-range
Ising chains at $0\leqslant\alpha\leqslant1$, in spite of the fact
that in our model the disorder is implemented in long-range interactions,
in a way quite different from the traditional method in the literature,
i.e., through random transverse field \cite{Canovi2011,Yao2014} or
random potential \cite{Oganesyan2007,Znidaric2008,Luitz2015,Baygan2015,Hauke2015,BarLev2015}.
Our results seems to rule out the possibility of interaction-assisted
delocalization at $1<\alpha<2$, as predicted by the resonant spin-pair-excitation
argument \cite{Yao2014,Burin2015a}. On the other hand, it is worth
noting that, our Ising model might be realized with the trapped ion
quantum emulator \cite{Smith2015} or randomly distributed Rydberg
atoms or polar molecules in lattices, and thus our results may stimulate
new directions for further experiments.

\section{Numerical Simulations}

\subsection{The model Hamiltonian}

We consider the following Ising model with disordered long-ranged
interactions, described by the model Hamiltonian with open boundary
condition, 
\begin{equation}
\mathcal{H}=J\sum_{1\leqslant i<j\leqslant L}\frac{\left(1+h_{i}h_{j}\right)}{\left|j-i\right|^{\alpha}}\sigma_{i}^{z}\sigma_{j}^{z}+B\sum_{i=1}^{L}\sigma_{i}^{x},\label{(H)}
\end{equation}
where $\sigma_{i}^{x}$ and $\sigma_{i}^{z}$ are the Pauli matrices
representing the $i$-th spin in the spin chain with $L$ spins. $B$
is a uniform transverse field and $J=1$ sets the energy unit of the
model. The exponent $\alpha$ characterizes the range of interactions,
where a larger $\alpha$ implies a shorter interaction range. $\{h_{i}\}$
is a set of $L$ independent dimensionless random variables, with
uniform distribution in the domain $[-W,W]$, where $W$ is defined
as the \emph{dimensionless} disorder strength\emph{.} Alternatively,
one may replace $h_{i}h_{j}$ with a single variable $h_{ij}$ that
distributes uniformly in some intervals; but the results should not
change qualitatively. Here, the disorder enters into the system entangling
with the long-range interaction in a non-trivial way. In the previous
work \cite{Hauke2015}, the renormalization group analysis reveals
that through renormalization, disorder on the transverse field induces
effective disorder on interactions. However, theoretical investigations
into models with disorder on interactions are relatively rare. Therefore,
a directly simulation of Eq. (\ref{(H)}) can to some extent provide
theoretical predictions for the behavior of system with quenched disorder
on \emph{interactions} and thereby testify the universal MBL physics
in an entirely new situation. In addition, the simulation also directly
addresses the following question: How does the interaction range influence
the MBL phase transition phenomenon? On the other hand, we anticipate
that the disorder on long-range interactions is experimentally realizable,
e.g. via programmable disorder with trapped $^{171}\textrm{Yb}^{+}$
ions \cite{Kim2011}. Polar molecules or Rydberg atoms \cite{Deng2016},
whose positions are randomly distributed in lattices, could also be
a candidate system to realize the proposed Ising Hamiltonian in Eq.
(\ref{(H)}). It should be noted that, the Hamiltonian Eq. (\ref{(H)})
has a $Z_{2}$ symmetry, i.e. the parity operator $\mathcal{P}=\prod_{1\leq i\leq L}\sigma_{i}^{x}$
commutes with the Hamiltonian. Throughout the paper, unless specified
otherwise, all numerical results are obtained from eigenstates with
a given parity of $\mathcal{P}=+1$, via exact diagonalization.

In our simulations, we find that, when $\alpha$ is tuned from $0.3$
to $3.0$, the system shows a drastic change in its global behavior.
Even though our numerical study of exact diagonalization suffers from
the small system size (up to $L=14$), our results on the two \emph{sensitive}
MBL indictors - the spectral statistics and half-chain entanglement
entropy - may contribute to clarify the debate in the proposed phase
diagram in the long-range Ising model (see Fig. \ref{PhaseDiagramPreviousWorks}c).

\subsection{The details of numerical calculations }

We have defined a dimensionless relative energy as, 
\begin{equation}
\epsilon=\frac{E-E_{\min}}{E_{\max}-E_{\min}},
\end{equation}
where $E_{\min}$ and $E_{\max}$ are the minimum and maximum energies
of the many-body system in a single realization of disorder configuration,
respectively. In this work, numerical results are obtained from the
$50$ eigenstates with their relative energy closest to $59/120$,
sampled over $1000$ different disorder configurations. \emph{B} is
always set to \emph{$0.6J$}, throughout the paper.

Usually, for simulations of models with long-range interactions, especially
for interactions decaying in the form of $R^{-\alpha}$, where $R=\left|i-j\right|$
is the distance between two spins of the system at positions $i$
and $j$, proper scaling factors - such as the Kac prescription \cite{Ioffe2010}
- are adapted to ensure that the Hamiltonian satisfies extensiveness
property. This is particularly important for the $\alpha<1$ case
since the average energy per spin will diverge as the system goes
to the thermodynamic limit (i.e., the size $L$ becomes infinitely
large) without such rescaling factor. However, an extra varying parameter
renders the analysis of numerical results less clear. To make things
even worse, different ways of rescaling might lead to quantitatively
different results, even though the qualitative behaviors of the system
may not be influenced. Therefore, in this work, no effort is made
to guarantee that the model Hamiltonian is extensive. Nevertheless,
the good agreement between our results and the predictions at $0\leqslant\alpha\leqslant1$
in a previous work using Kac prescription by Hauke and Heyl \cite{Hauke2015}
indicates that the qualitative physics is not changed with or without
the Kac prescription.

\section{Results and discussions}

\subsection{Spectral statistics}

The energy spectral statistics or the statistical energy gap distribution
is a sensitive indicator of the MBL phase transition \cite{Oganesyan2007}.
According to the random matrix theory, because of the overlapping
of local freedoms between different eigenstates in a thermalized phase,
gaps between adjacent energy levels should obey a Wigner-Dyson distribution,
the so-called Gaussian orthogonal ensemble (GOE) in our case. Due
to the lack of such overlappings in a localized phase, these gaps
instead should follow a Poisson distribution. It is convenient to
characterize these two different distributions using the averaged
ratio of successive gaps, $\left\langle r\right\rangle $ defined
by \cite{Oganesyan2007} 
\begin{equation}
\left\langle r\right\rangle =\left\langle \frac{\min\left\{ \delta_{n+1},\delta_{n}\right\} }{\max\left\{ \delta_{n+1},\delta_{n}\right\} }\right\rangle ,
\end{equation}
where $\delta_{n}=E_{n+1}-E_{n}$ and $\left\langle \cdots\right\rangle $
denotes an average over some eigenstates calculated within a given
disorder realization and over all 1000 different disorder configurations
simulated. The GOE distribution has $\left\langle r\right\rangle \approx0.5307(1)$,
while the Poisson distribution results in $\left\langle r\right\rangle =2\ln2-1\approx0.3863$
\cite{Atas2013}. In the following, these two featured values are
denoted by $r_{G}$ and $r_{P}$, respectively.

\begin{figure}
\centering{}\includegraphics[width=0.48\textwidth]{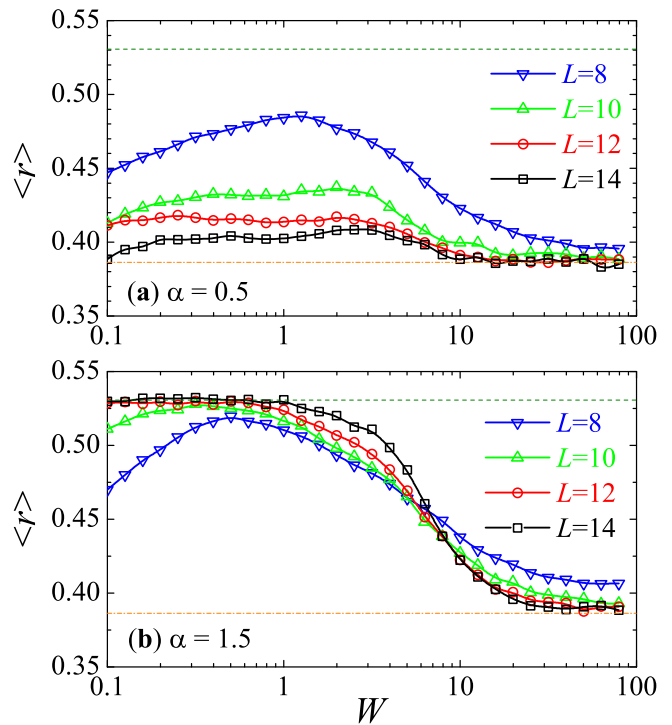} \caption{(color online). The averaged ratio of the adjacent energy spacings
$\left\langle r\right\rangle $ of the Hamiltonian Eq. (\ref{(H)})
at $\alpha=0.5$ (a) and $\alpha=1.5$ (b) for four different lengths
$L=8$, $10$, $12$ and $14$. The dashed and dot-dashed lines indicate
$r_{G}\approx0.5307(1)$ and $r_{P}\approx0.3863$, respectively.
\label{R1}}
\end{figure}

Typical disorder dependence of the averaged ratio $\left\langle r\right\rangle $,
as illustrated in Fig. \ref{R1}, reveals that there is a drastic
change when the exponent of the interaction range $\alpha$ is tuned
from $0.3$ to $3.0$. For example, at $\alpha=0.5$ {[}see Fig. \ref{R1}(a){]},
the ratio decreases monotonically with increasing system size $L$.
It is always smaller than $r_{G}$ anticipated for a thermalized state.
In contrast, at $\alpha=1.5$ {[}see Fig. \ref{R1}(b){]}, with increasing
$L$ the ratio increases at small disorder but decreases at sufficiently
large disorder strength. This leads to an apparent crossing point
$W_{c}$, which locates the critical disorder strength for the MBL
phase transition \cite{Oganesyan2007}. For the case with $\alpha=1.5$
in Fig. \ref{R1}(b), we find that $W_{c}\approx6.5$. It is worth
emphasizing that all the lines in Fig. \ref{R1} approach the Poisson
$r_{P}$ limit at sufficiently large disorder strength, implying that
the system is always in MBL phase for strong enough disorder.

\begin{figure}
\centering{}\includegraphics[width=0.48\textwidth]{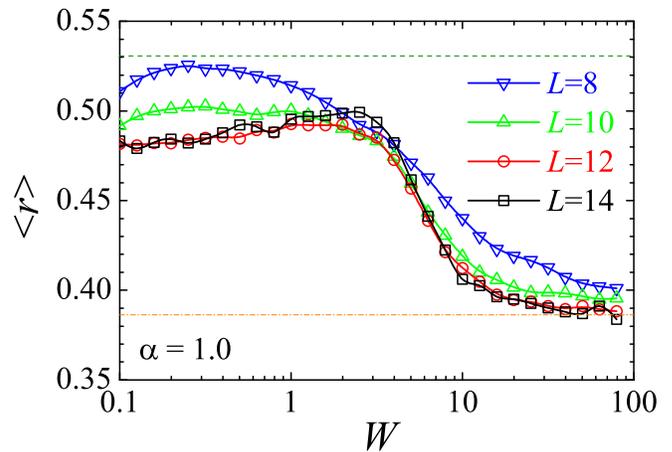} \caption{(color online). The averaged ratio of the adjacent energy spacings
$\left\langle r\right\rangle $ of the Hamiltonian Eq. (\ref{(H)})
at $\alpha=1$. \label{R2}}
\end{figure}

By carefully examining the disorder dependence of $\left\langle r\right\rangle $
at various exponent of the interaction range, it turns out that $\alpha_{c}=1$
is very likely to be the threshold exponent, above and below which
the system exhibits entirely different responses at weak disorder.
This is evident in Fig. \ref{R2} showing the averaged ratio at $\alpha=1$
in four curves for four different system sizes. We notice that there
is no crossing point of these curves. In fact, all these four curves
are nearly tangent with each other at $W\approx3.5$. Therefore, it
is reasonable to assume the following phase structure: 
\begin{itemize}
\item For $\alpha<1$, the system is mostly localized for nonzero disorder
strength. But, there are probably two kinds of localized phases, separated
by a quantum critical region; and 
\item For $\alpha>1$, in the small limit of disorder strength, the system
is in the thermalized phase that satisfies ETH, while in the strong
disorder limit, the system is in the MBL localized phase. These two
phases are connected via an MBL phase transition.
\end{itemize}
These phase diagram structure are confirmed by studying another popular
indicater: quantum entanglement entropy.

\subsection{Entanglement Entropy}

The entanglement entropy $S_{E}$ has been widely used in the literature
\cite{Zhao2015,Pal2010,Bardarson2012,Grover2014,Li2015,Serbyn2013a,Baygan2015}
to characterize the MBL. The meaning and implication of this indicator
is clear. One of the most important features of MBL (as is indicated
by its name) is localization. Local freedoms in an MBL eigenstate
are no longer entangled with freedoms that are far away in real space.
Thus, the entanglement entropy of a subsystem should follow an area
law deeply in MBL phase rather than the volume law found in equilibrated
states. Specifically, for our Hamiltonian, the half-chain entanglement
entropy deeply in MBL phase should be independent of the length of
the system and very close to $\ln2$. The origin of the $\ln2$ is
due to the $Z_{2}$ parity symmetry. A detailed explanation is given
in Appendix A.

The effective temperature for an eigenstate with energy $E_{eig}$
can be defined as: 
\begin{equation}
E_{eig}=\frac{\textrm{Tr}\left(\mathcal{H}e^{-\beta\mathcal{H}}\right)}{\textrm{Tr}\left(e^{-\beta\mathcal{H}}\right)},\label{temperature}
\end{equation}
where $\beta=1/k_{B}T$ with $k_{B}$ being the Boltzmann constant
and $T$ effective temperature.

In this work, we are focusing on eigenstates with relative energy
close to $59/120$, whose effective temperature should therefore be
very close to infinity. Thus, in thermal phases, the half-chain entanglement
entropy should be close to $(L\ln2-1)/2$, the classical entropy of
the half chain at infinite temperature \cite{Page1993}. Also, as
is implied by the Eq. (\ref{temperature}), the relative position
of an eigenstate in the energy spectrum can drastically influence
its effective temperature. Under the assumption that it is the portion
of the averaged spectrum rather than the absolute number of the eigenstates
that matters for the structure of entanglement entropy, we calculate
the entanglement entropy of different numbers of eigenstates for Hamiltonians
with different spin numbers to ensure that the averaged eigenstates
occupy a constant portion of the whole spectrum. Specifically, for
chains with 8, 10 and 12 spins, the entanglement entropy $S_{E}$
and uncertainty $\delta S_{E}$ are calculated by using $3$, 12 and
48 eigenstates with relative energy closest to $59/120$ respectively.
In priciple, we should have used $192$ eigenstates for a chain with
$14$ spins, but the calculation turned out to be too heavy and time-consuming.
Thus, results of chain with $14$ spins are calculated using only
$50$ eigenstates with relative energy closed to 59/120. Nevertheless,
the result of the chain with $14$ spins still shows a similar pattern
as those shorter chains, indicating our choice is acceptable.

\begin{figure}
\centering{}\includegraphics[width=0.48\textwidth]{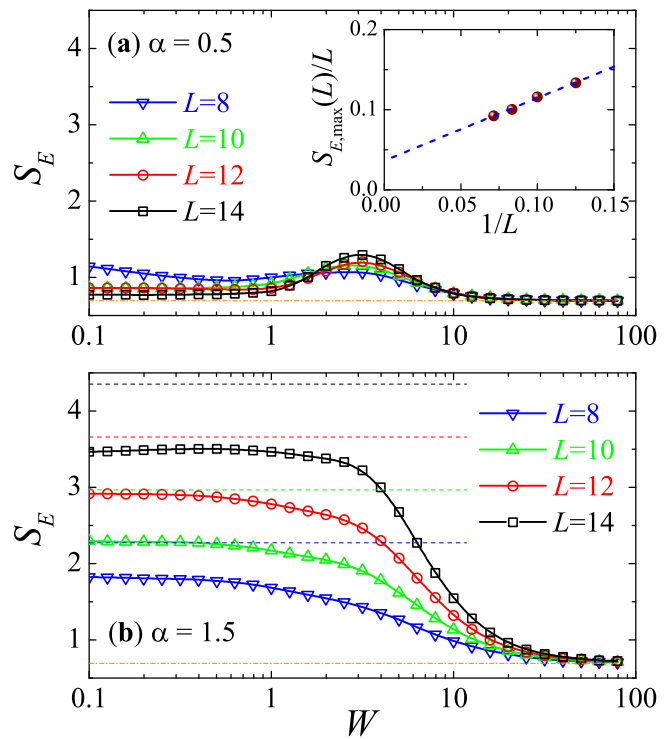} \caption{(color online). The half-chain entanglement entropy $S_{E}$ of the
Hamiltonian Eq. (\ref{(H)}) at $\alpha=0.5$ (a) and $\alpha=1.5$
(b) for four different lengths $L=8$, $10$, $12$ and $14$. The
thin dot-dashed lines in both subplots indicate the $\ln2$ entropy
anticipated in the deep MBL phase and the dashed lines in (b) indicate
the expected classical entropy $(L\ln2-1)/2$ in thermalized phase.
The inset in (a) shows the maximum entropy per spin (found at the
intermediate disorder strength) as a function of $1/L$. \label{EE}}
\end{figure}

Our numerical simulations shows that the exponent $\alpha$ of the
interaction range drastically changes the behavior of the entanglement
entropy. As can be seen clearly in Fig. \ref{EE}(a), in the $\alpha=0.5$
case, the half-chain entanglement entropy at both weak and strong
disorder strengths is largely independent on the length of the system
and always lies very close to $\ln2$. This observation seems to be
consistent with the earlier prediction by Hauke and Heyl \cite{Hauke2015},
which states that for $\alpha<1$ the disordered Ising chain is always
localized for any finite disorder strength. However, at the intermediate
disorder strength, the entanglement entropy also shows a pronounced
peak structure, with a maximum entropy that increases with increasing
system size. A close examination of the size dependence is given in
the inset of Fig. \ref{EE}(a). We find that the maximum entropy per
spin, $S_{E}(L)/L$, scales to a finite value ($\approx0.04\ll0.5\ln2$)
in the thermodynamic limit of $L\rightarrow\infty$. Thus, the maximum
entropy seems to follow a \emph{weak} volume law, instead of the area
law obeyed in the localized phase. Therefore, there is a possibility
that, at $\alpha=0.5$, two different MBL phases may exist at weak
and strong disorder strengths, respectively. They are separated by
a quantum critical region at intermediate disorder strength, in which
the entanglement entropy follows a weak volume law. At $\alpha=1.5$,
Fig. \ref{EE}(b) shows that the entanglement entropy has a linear
dependence on the system size at weak disorder strength and approaches
$\ln2$ at strong disorder strength. This behavior was found earlier
in a disordered spin chain with nearest-neighbor interactions \cite{Kjall2014,Luitz2015},
where a phase transition from a thermalized state to the MBL phase
is now well established. 

\begin{figure}
\centering{}\includegraphics[width=0.48\textwidth]{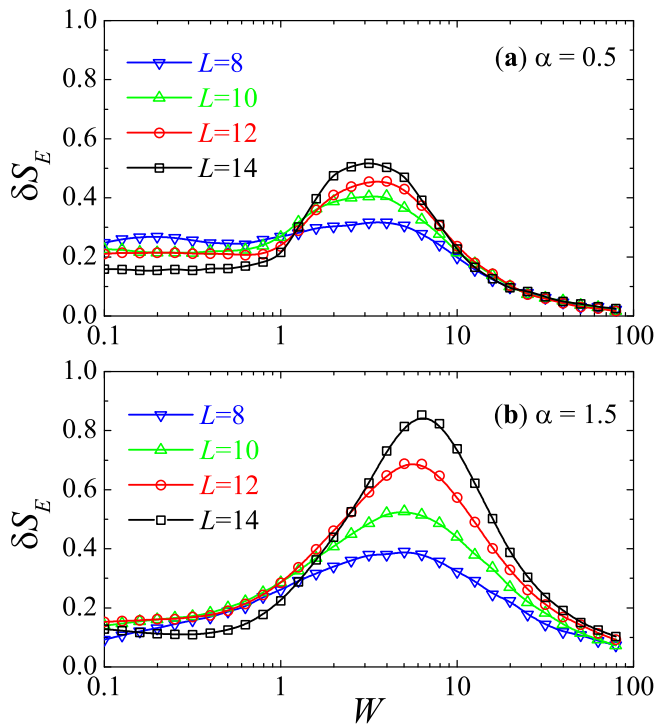} \caption{(color online). The uncertainty of the half-chain entanglement entropy
$\delta S_{E}$ of the Hamiltonian Eq. (\ref{(H)}) at $\alpha=0.5$
(a) and $\alpha=1.5$ (b) for four different lengths $L=8$, $10$,
$12$ and $14$. \label{dE} }
\end{figure}

The above picture is further supported by the behavior of the uncertainty
of the entanglement entropy $\delta S_{E}$, as shown in Fig. \ref{dE}.
For the $\alpha=1.5$ case in Fig. \ref{dE}(b), in the small disorder
limit, the uncertainty is small, consistent with the fact that the
system is in a thermal state and hence the half-chain entanglement
entropy should approach the constant classical limit. In the strong
disorder limit, the uncertainty also gradually decreases to zero,
in agreement with the fact that the system is localized and thus its
half-chain entanglement entropy approaches the constant value of $\ln2$.
At intermediate disorder strength, the half-chain entanglement entropy
of the eigenstates can be either close to $\ln2$ or extensively large,
leading to a large uncertainty, which can be viewed as an excellent
signature of the MBL phase transition.

For the $\alpha=0.5$ case in Fig. \ref{dE}(a), on the other hand,
the system is mostly in the exotic localized states, as we mentioned
earlier. It becomes non-trivial to predict the behavior of the uncertainty
of the entanglement entropy. Nevertheless, because the entanglement
entropy itself is small and nearly independent on the length of the
chain, we anticipate a small uncertainty, which is indeed seen in
Fig. \ref{dE}(a) at both weak and strong disorder strengths. The
peak structure of the uncertainty at intermediate disorder strength
also seems to be consistent with the weak volume law of the entanglement
entropy discussed earlier. Similar to the $\alpha=1.5$ case, we may
regard it as the indication of a possible phase transition between
two MBL phases.

To confirm that the proposed phase diagram structure is qualitatively
correct in the thermaldynamic limit, we now turn to perform a finite
size scaling.

\subsection{Finite size scaling}

The data we used for performing the finite size scaling analysis are
the half-chain entanglement entropy $S_{E}$ and the related uncertainty
$\delta S_{E}$. By suitably defining a scaled disorder strength and
scaled $S_{E}$ and $\delta S_{E}$, we anticipate that all the data
with different system sizes will collapse onto a single curve near
possible quantum phase transition. Through data collapse, one may
be able to extract useful information about the phase transition and
determine the critical disorder strength. Unfortunately, currently
we do not have a well-established theory to provide us with the reliable
scaling form yet. Thus, in this work, we adapt the most popular scaling
form used in the previous studies \cite{Luitz2015},

\begin{equation}
Q\left(L,W\right)=g\left(L\right)f\left[\left(W-W_{c}\right)L^{1/\nu}\right],
\end{equation}
where $Q$ stands for $S_{E}$ or $\delta S_{E}$ and $g(L)=[(L-2)\ln2-1]/2$
is the difference between the two limiting entropies in the thermalized
phase and in the MBL phase. It is used as a pre-factor to rescale
the entanglement entropy and the related uncertainty. $\nu$ is known
as the critical exponent and $W_{c}$ is the critical disorder strength.
Both of them are treated as the variational parameters, in order to
scale the data with different $L$ onto a single scaling curve $f(x)$.

\begin{figure}
\centering{}\includegraphics[width=0.48\textwidth]{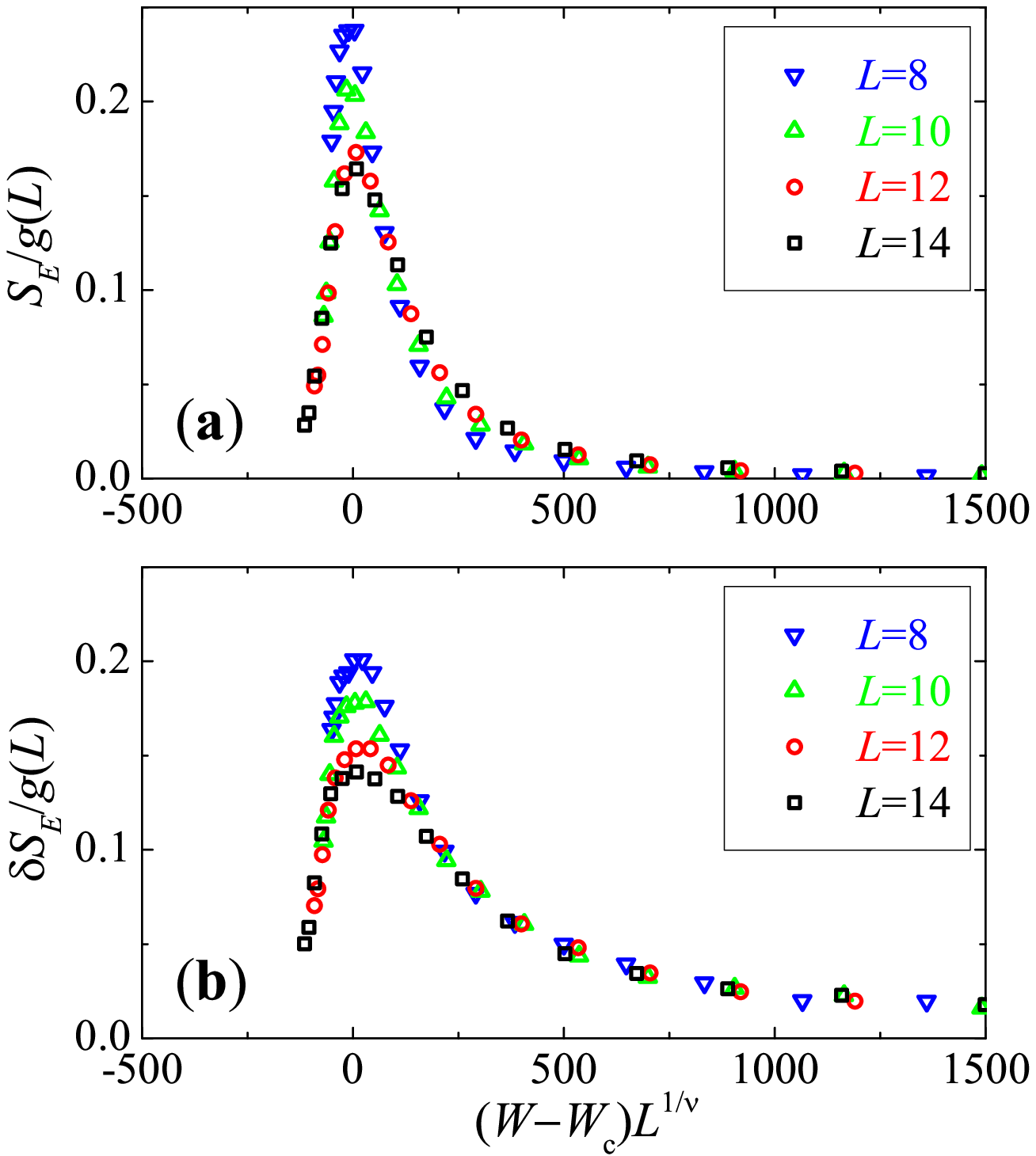} \caption{(color online). The entanglement entropy and its uncertainty, in units
of $g(L)=[(L-2)\ln2-1]/2$, as a function of the scaled disorder strength
$(W-W_{c})L^{1/\nu}$ at $\alpha=0.5$. \label{ScalingAlpha05}}
\end{figure}

\begin{figure}
\centering{}\includegraphics[width=0.48\textwidth]{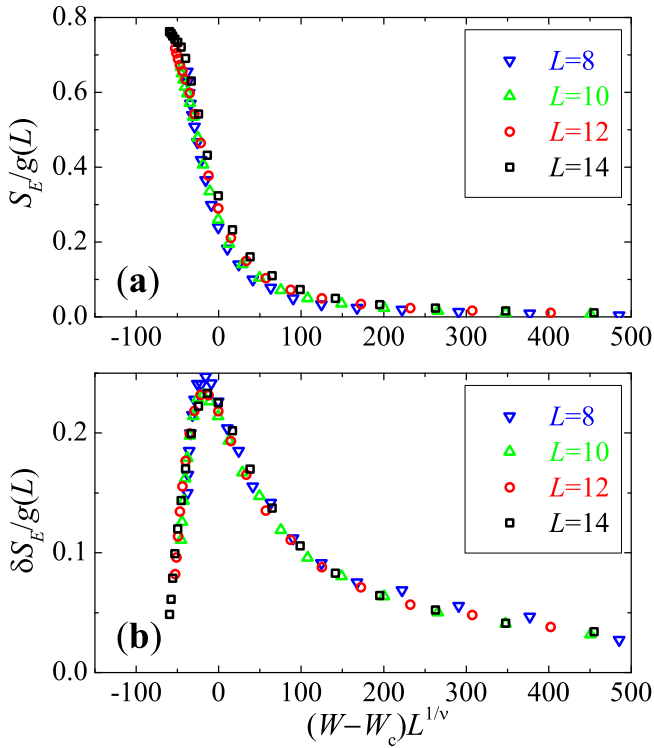} \caption{(color online). The entanglement entropy and its uncertainty, in units
of $g(L)$, as a function of the scaled disorder strength $(W-W_{c})L^{1/\nu}$
at $\alpha=1.5$. \label{ScalingAlpha15}}
\end{figure}

In Fig. \ref{ScalingAlpha05} and Fig. \ref{ScalingAlpha15}, we present
the scaled entanglement entropy and its uncertainty at $\alpha=0.5$
and $\alpha=1.5$, respectively. By suitably adjusting the two parameters
$\nu$ and $W_{c}$, the originally scattered data of $S_{E}$ (in
Fig. \ref{EE}) or $\delta S_{E}$ (in Fig. \ref{dE}) indeed collapse
onto a single curve, as one may anticipate. The data collapse at $\alpha=1.5$
is particularly satisfactory, confirming the existence of a MBL phase
transition. This conclusion disagrees with a previous prediction from
the resonant spin-pair excitations arguement that a disordered transverse-field
Ising chain should always be delocalized for $1<\alpha<2$ \cite{Burin2015a}. 

\begin{figure}
\centering{}\includegraphics[width=0.48\textwidth]{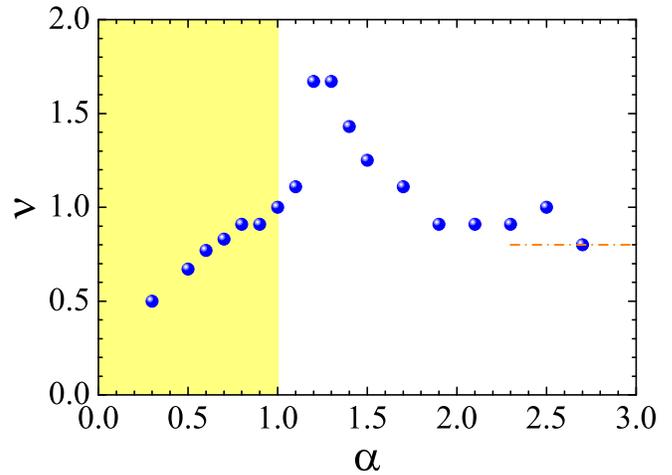} \caption{(color online). The critical exponent $\nu$ of the Hamiltonian Eq.
(\ref{(H)}) as a function of the exponent of the interaction range
$\alpha$. The dot-dashed line indicates the result of a disordered
Heisenberg chain with nearest-neighbor interactions, $\nu=0.80\pm0.04$
\cite{Luitz2015}. \label{CriticalExponent}}
\end{figure}

\subsection{Phase diagram}

Finally, we performed the finite size scaling at different exponent
of the interaction range $\alpha.$ The resulting critical disorder
strengths and critical exponents are shown in the phase diagram Fig.
\ref{PhaseDiagramL14} (circles) and Fig. \ref{CriticalExponent},
respectively. To obtain the critical disorder strength, we have also
used the averaged gap ratio $\left\langle r\right\rangle $ and extract
the position of the cross point, see for example, Fig. \ref{R1}(b).
This is another useful way valid down to $\alpha=1$. In the phase
diagram Fig. \ref{PhaseDiagramL14}, the critical disorder strengths
determined using $\left\langle r\right\rangle $ are plotted by diamond
symbols. In general, the phase boundary determined from the two methods
coincides well. This suggests that for our model Hamiltonian Eq. (\ref{(H)}),
the averaged ratio of successive gaps and the finite-size scaling
analysis of the entanglement entropy as well as its uncertainty are
the equivalent methods to pinpoint the phase transition. 

For the critical exponent, as shown in Fig. \ref{CriticalExponent},
there is a peak around $\alpha=1$. This is a strong indication that
$\alpha_{c}=1$ is a threshold exponent, which separates the phase
diagram Fig. \ref{PhaseDiagramL14} into two parts. When $\alpha$
is small, the critical exponent trends to $0.5$. When $\alpha$ is
large, the critical exponent seems to saturate to $0.9$, a value
that is close to the critical exponent obtained for a disordered Heisenberg
chain with nearest-neighbor interactions at the same infinite temperature,
i.e., $\nu=0.80\pm0.04$ \cite{Luitz2015}. This is understandable,
since, as the exponent of the interaction range $\alpha$ goes to
infinity, the long-range interaction in our model Hamiltonian Eq.
(\ref{(H)}) is naturally reduced to a short-range interaction.

\section{Summary}

In conclusions, we have investigated a one-dimensional Ising spin
model with random long-range interactions, a system that may be relevant
to trapped ions, Rydberg atoms and polar molecules in cold-atom experiments.
By systematically studying the two many-body localization indicators
such as the averaged ratio of successive energy gaps and the entanglement
entropy (and its uncertainty), and performing the finite size scaling,
we have found that the system always experiences a many-body localization
phase transition at sufficiently large disorder strength. To some
extent, this is a surprising result, as the previous theoretical investigation
suggested a complete delocalization at the exponent of the interaction
range $1<\alpha<2$, due to the picture of resonant spin-pair excitations
\cite{Burin2015a}. A phase diagram has been determined (see Fig.
\ref{PhaseDiagramL14}), as functions of the disorder strength $W$
and the interaction exponent $\alpha$. We have determined the phase
boundary and have found that $\alpha_{c}=1$ is a threshold interaction
exponent. For $\alpha>1$, the system undergoes a thermal-MBL phase
transition with increasing disorder strength; while for $\alpha<1$,
the system is mostly many-body localized, a result in agreement with
a previous finding \cite{Hauke2015}. There could be two different
localized phases, separated by a quantum critical region, whose properties
are yet to be understood.
\begin{acknowledgments}
This work was supported by the Australian Research Council (ARC) Future
Fellowship grants (Grant Nos. FT140100003 and FT130100815) and Discovery
Projects (Grant Nos. DP140100637 and DP140103231). All the numerical
calculations were performed using Swinburne new HPC resources (Green
II) at Swinburne University of Technology.
\end{acknowledgments}

\appendix

\section{The origin of $\ln2$ entanglement entropy in the deep MBL phase}

As mentioned and demonstrated in Section III, deeply in the MBL phase
the half chain entanglement entropy approaches $\ln2$ in the large
disorder limit. This $\ln2$ entropy originates from the $Z_{2}$
parity symmetry of the Hamiltonian. To prove this, let us consider
the half chain entanglement entropy of other two Hamiltonians under
similar parameters and conditions, 
\begin{equation}
\mathcal{H}_{1}=J\sum_{1\leqslant i<j\leqslant L}\frac{1}{|j-i|^{\alpha}}\sigma_{i}^{z}\sigma_{j}^{z}+J\sum_{i=1}^{L}h_{i}\sigma_{i}^{x},\label{H1}
\end{equation}
\begin{equation}
\mathcal{H}_{2}=J\sum_{1\leqslant i<j\leqslant L}\frac{1+h_{i}h_{j}}{|j-i|^{\alpha}}\sigma_{i}^{z}\sigma_{j}^{z}+B\sum_{i=1}^{L}\sigma_{i}^{x}+C\sum_{i=1}^{L}\sigma_{i}^{z},
\end{equation}
where ${h_{i}}$ is the dimensionless random variable drawn from the
uniform distribution in the domain $[-W,W]$. The Hamiltonian $\mathcal{H}_{1}$
is simply the random transverse-field Ising model considered earlier
by Hauke and Heyl \cite{Hauke2015} and by Burin \cite{Burin2015a}.
In the Hamiltonian $\mathcal{H}_{2}$, a longitudinal field is applied.
Both $\mathcal{H}_{1}$ and $\mathcal{H}_{2}$ break the $Z_{2}$
parity symmetry. To be specific, we consider the case with $\alpha=1.5$
and $L=8$.

The half-chain entanglement entropies of three eigenstates with relative
energy closest to $1/2$ (for $\mathcal{H}_{1}$) and $59/120$ (for
$\mathcal{H}_{2}$) are calculated and averaged over $1000$ different
disorder configurations. To ensure that the comparison is carried
out under similar conditions, in the Hamiltonian $\mathcal{H}_{2}$,
$B$ is set to $0.6J$ as before, while $C$ is taken to be $0.01J$.
We also consider the Hamiltonian Eq. (\ref{(H)}) \emph{without} imposing
the parity constraint of $\mathcal{P}=+1$.

\begin{figure}
\centering{}\includegraphics[width=0.48\textwidth]{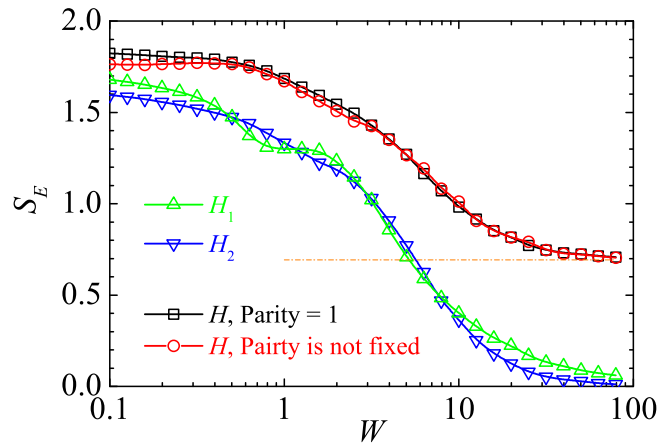} \caption{The half chain entanglement entropy $S_{E}$ of different model Hamiltonians
at $\alpha=1.5$ and $L=8$. The thin dot-dashed line indicates the
$\ln2$ entanglement entropy observed in the deep MBL regime for the
model Hamiltonian $\mathcal{H}$. \label{OriginEntropyln2}}
\end{figure}

In Fig. \ref{OriginEntropyln2}, we compare the half-chain entanglement
entropies of four different situations: $\mathcal{H}_{1}$ (up triangles),
$\mathcal{H}_{2}$ (down triangles), $\mathcal{H}$ with the $Z_{2}$
parity $\mathcal{P}=+1$ (squares), and $\mathcal{H}$ without the
parity constraint (circles). It is clear that, as soon as the $Z_{2}$
parity symmetry is broken, no matter it is destroyed by a disordered
transverse field or by a uniform longitudinal field, deeply in the
localized phase, the entanglement entropy goes to zero. On the other
hand, the parity constraint implemented in our calculations (as in
the main text) has essentially no quantitative influence on the entanglement
entropy. This is understandable since the model Hamiltonian respects
the $Z_{2}$ parity symmetry and then, in principle, the eigenstates
solved by exact diagonalization would have a deterministic parity.
With nearly the same energy, these eigenstates (having either $\mathcal{P}=+1$
or $\mathcal{P}=-1$) would have nearly the same entanglement entropy.

However, there are two main reasons that renders the implementation
of the parity constraint preferable. Firstly, when the disorder strength
is very large, all the eigenstates of the Hamiltonian $\mathcal{H}$
become nearly doubly degenerate, due to the $Z_{2}$ parity symmetry.
Our results are all calculated using Matlab via its \emph{'eigs}'
function. Due to the limitation of its inherent algorithm, this function
can not precisely distinguish two nearly degenerate eigenstates. If
a parity is not settled from the beginning, it will produce non-physical
entanglement entropy due to the parity mixing. On the other hand,
the use of a deterministic parity can reduce the dimension of the
Hilbert space by half. This is highly preferable in numerical simulations.

\bibliographystyle{apsrev4-1}
\bibliography{References}

\end{document}